# Correlation Effects on Magnetic Structure and Lattice Dynamics of LaMn$_7$O$_{12}$: A First-Principles Study

Haeyoon Jung[1], Indukuru Ramesh Reddy[2], Bongjae Kim[2], Jiyeon Kim[1,3*], Sooran Kim[1,4*]

[1]Department of Physics Education, Kyungpook National University, Daegu 41566, South Korea

[2]Department of Physics, Kyungpook National University, Daegu 41566, South Korea

[3]The Center for High Energy Physics, Kyungpook National University, Daegu 41566, South Korea

[4]KNU G-LAMP Project Group, KNU Institute of Basic Sciences, Kyungpook National University, Daegu, 41566, Korea.

*Corresponding authors: mygromit@gmail.com, sooran@knu.ac.kr

**Abstract**

LaMn$_7$O$_{12}$, a quadruple perovskite oxide (AA′$_3$B$_4$O$_{12}$-type), has attracted attention for its notable bifunctional activity in oxygen evolution and reduction reactions. Here, we systematically investigate the magnetic phase diagram and lattice dynamics of LaMn$_7$O$_{12}$ using two density functional theory plus Hubbard $U$ (DFT + $U$) approaches: the spin-density and the charge-only-density formalism. Phase diagram analysis as a function of $U$ and $J$ shows that both methods stabilize the experimentally observed antiferromagnetic (AFM) configuration (C-type AFM at the B-site and ferrimagnetic structure at the A′-site Mn ions) at $U$ = 3.5 eV and $J$ = 0.8 eV. These $U$ and $J$ values are consistent with those obtained from the constrained random phase approximation. Furthermore, we observe the dynamical stability of the AFM phase through phonon dispersion curves and analyze the Raman-active phonon modes. These results highlight the critical role of appropriate $U$ and $J$ parameters in accurately describing the properties of LaMn$_7$O$_{12}$.

## Introduction

The efficiency of energy conversion systems such as fuel cells, metal-air batteries, and water electrolysis devices strongly depends on the key electrochemical processes: the oxygen evolution reaction (OER) and the oxygen reduction reaction (ORR) [1,2]. Currently, noble metal catalysts such as $RuO_2$ and $IrO_2$ are widely used for the OER, while Pt serves as the standard catalyst for the ORR [3–5]. Although these catalysts exhibit excellent activity, their high cost and limited availability restrict their application on a large scale. To address these limitations, recent research has focused on developing earth-abundant alternatives that can efficiently catalyze both OER and ORR on a single electrode, commonly referred to as bifunctional catalysts [6,7].

Perovskite oxides have been regarded as promising alternatives to noble-metal catalysts, owing to their structural tunability, electronic versatility, and use of earth-abundant elements [8–11]. For example, $Ba_{0.5}Sr_{0.5}Co_{0.8}Fe_{0.2}O_{3-\delta}$ exhibits OER activity, which is an order of magnitude higher than that of $IrO_2$, with its near-unity $e_g$ occupancy [11]. Beyond these conventional $ABO_3$-type structures, quadruple perovskites with the formula $AA'_3B_4O_{12}$ have also been extensively investigated, as their ordered cation sublattices and complex electronic configurations provide additional degrees of freedom for optimizing catalytic performance [12–14].

Among various quadruple perovskites, $LaMn_7O_{12}$ has recently emerged as an attractive bifunctional electrocatalyst. It exhibits enhanced OER and ORR activities, which have been attributed to its dual-site reaction mechanism and to the durability arising from a robust covalent bonding network [12–14]. Alongside its electrocatalytic functionality, $LaMn_7O_{12}$ has also been studied as a model system for exploring correlated electronic and structural phenomena. It crystallizes in the monoclinic space group $I2/m$ phase below 653 K, exhibiting the Jahn-Teller (J-T) distortions at the B-site $Mn^{3+}$ ions and insulating behavior at ambient conditions [15–17]. Previous experimental studies have reported intriguing properties, including temperature- and pressure-induced structural phase transitions to cubic space group $Im\bar{3}$ symmetry with the suppression of the J-T distortions [15,17], complex magnetic transitions and orderings in A′ and B Mn sublattices at ~20 K and ~80 K [16,18,19], and ferroelectricity arising from canted AFM ordering of the B-site Mn [20]. Collectively, these findings highlight the potential of $LaMn_7O_{12}$ for understanding the interplay among spin, charge, orbital, and lattice degrees of freedom in transition-metal oxides.

Despite its interesting experimental observations, theoretical studies on $LaMn_7O_{12}$ remain relatively limited. Experimentally, two different magnetic configurations have been claimed to be the ground ones: C-type AFM at the B-site Mn ions with A-type AFM at the A′-site in 2009 [16] and C-type AFM at the B-site with ferrimagnetic structure at the A′-site Mn ions in 2018 [19]. Most of the computational studies have investigated the electronic structures of $LaMn_7O_{12}$ with simple FM or AFM orderings using the DFT+$U$ scheme [12,14,21,22]. For example, Liu *et al.* explored the ground states and density of states by considering four different magnetic configurations [21]. However, to the best of our knowledge, none of the computational studies have considered the recently reported experimental magnetic structure [19]. Coulomb interaction parameters play an essential role in accurately describing the electronic structures and magnetic configurations, which are important for predicting catalytic activity [21–23]. Thus, it is worth investigating the magnetic ground states and properties of $LaMn_7O_{12}$ by considering experimental magnetic structures and various $U$ parameters.

In this work, we systematically investigate the dependence of the magnetic ground state of $LaMn_7O_{12}$ on the Coulomb parameter $U$ and the Hund coupling $J$ by employing the

Liechtenstein DFT+$U$ formalism [24], which treats them as independent parameters. We also explore the two different DFT+$U$ implementations: the spin-density functional theory plus $U$ (SDFT+$U$) and charge-density functional theory plus $U$ (CDFT+$U$). We construct four magnetic configurations: one ferromagnetic (FM) and three antiferromagnetic (AFM) types, including two experimentally reported magnetic structures [16,19]. Their relative stabilities are compared through total energy calculations as functions of $U$ and $J$. From this analysis, we identify appropriate $U$ and $J$ parameters that reproduce the experimentally observed magnetic behavior. We further investigate the electronic and phonon structures using the identified $U$ and $J$ values. The dynamical stability and Raman modes of LaMn$_7$O$_{12}$ are analyzed from phonon calculations and compared with previous experiments.

**Computational Methods**

DFT calculations were carried out using the Vienna Ab *initio* Simulation Package (VASP). The calculations employed the generalized gradient approximation (GGA) with the Perdew-Burke-Ernzerhof (PBE) functional, combined with an on-site Coulomb interaction correction (PBE+$U$) for the Mn $d$ orbitals [25]. We further investigated two different types of the rotationally invariant DFT + $U$ schemes proposed by Liechtenstein *et al.* [24]: one considers the exchange-correlation functional dependent on spin density (SDFT+$U$), which is implemented in VASP using LDAUTYPE = 1, and the other is based on a charge-only-density dependent exchange correlation functional (CDFT+$U$), implemented using LDAUTYPE = 4 [26]. Unless otherwise specified, we used the experimental crystal structure of LaMn$_7$O$_{12}$ in a monoclinic space group *I*2/*m* whose lattice parameters of $a$ = 7.509 Å, $b$ = 7.349 Å, $c$ = 7.504 Å, and $\beta$ = 91.354° [16] to exclude contributions from the structural relaxation. The $k$-point sampling was 6×6×6, which corresponds to the $k$-point density of about 5000/atom.

The $U$ and $J$ values for the Mn-$d$ manifold were calculated using the constrained random-phase approximation (cRPA) with the projector method [27], as implemented in VASP. The Coulomb interaction parameters derived from this method have been demonstrated to provide a better description of the electronic structure [28–30]. Comprehensive theoretical formulations and technical details of the cRPA method can be found in Refs. [27,31–35]. For the cRPA calculations, a 6×6×6 $k$-mesh was used, and 324 bands were employed. A large plane-wave cutoff energy of 650 eV was used for self-consistent calculation, while 433 eV was used for dielectric function calculations. The Wannier functions were obtained using the WANNIER90 [36] with VASP2WANNIER [37] interface, employing default disentanglement and Wannierization parameters [29]. To construct the Wannier functions, we included all Mn-$d$ and O-$p$ bands within a disentanglement energy window of [-6.89, 3.62] eV (see Fig. S1).

We employed the PHONOPY software package to investigate the lattice dynamics and evaluate the structural stability of LaMn$_7$O$_{12}$ [38]. A finite-displacement method was applied to supercells to obtain the force constants and dynamic matrices. For phonon calculations, a 2×2×2 supercell and 3×3×3 $k$-point sampling were used, and the lattice parameters and atomic positions were fully relaxed until the forces converged below 0.01 eV/Å.

**Results and Discussion**

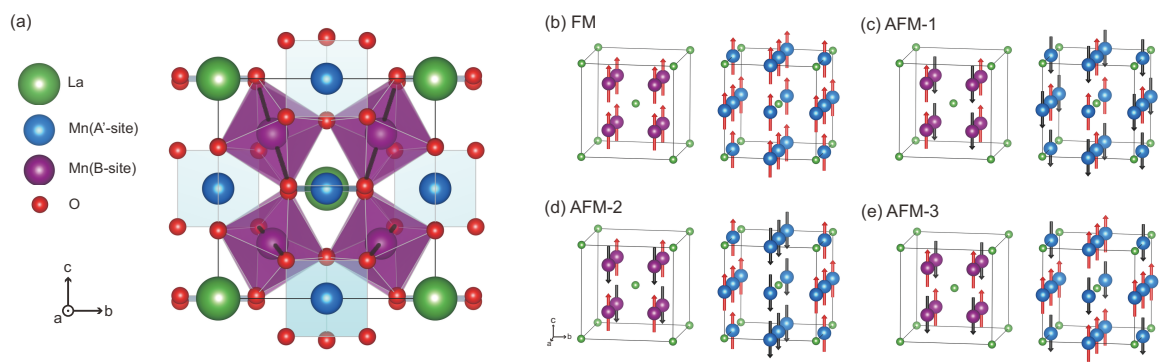

**Fig. 1** (a) Crystal structure of LaMn$_7$O$_{12}$. Green, blue, purple, and red circles represent La, A′-site Mn, B-site Mn, and O atoms, respectively. Black lines indicate elongated Mn-O bonds due to the Jahn-Teller distortion. (b-e) Magnetic structures: (b) FM, (c) AFM with G-type ordering, (d) A-type ordering, and (e) ferrimagnetic ordering at the A′-site Mn ions. All AFM configurations share C-type ordering in the B-site sublattice. To highlight the magnetic sites, La atoms are shown as small green spheres, while oxygen atoms are omitted.

LaMn$_7$O$_{12}$ adopts a quadruple perovskite structure of the AA′$_3$B$_4$O$_{12}$ type, where both the A′ and B sites are occupied by Mn ions as shown in Fig. 1(a) [14,16]. The A′-site Mn ions are coordinated in a square-planar arrangement with oxygen, while the B-site Mn ions are from MnO$_6$ octahedra [16]. The B-site Mn$^{3+}$ ions exhibit the J-T distortions, and the elongated Mn–O bonds are indicated by black lines in Fig. 1(a). The trivalent La$^{3+}$ ions maintain charge neutrality by stabilizing Mn ions in the 3+ oxidation state, resulting in a single-valent Mn$^{3+}$ system [16,18].

Based on this structure, four magnetic configurations were constructed: one ferromagnetic (FM) and three antiferromagnetic (AFM) configurations, as illustrated in Fig. 1(b–e). The AFM-1 configuration was reported as the most stable magnetic structure in a previous DFT study [21]. The AFM-2 and AFM-3 correspond to experimentally observed magnetic orderings reported in 2009 and 2018, respectively [16,19].

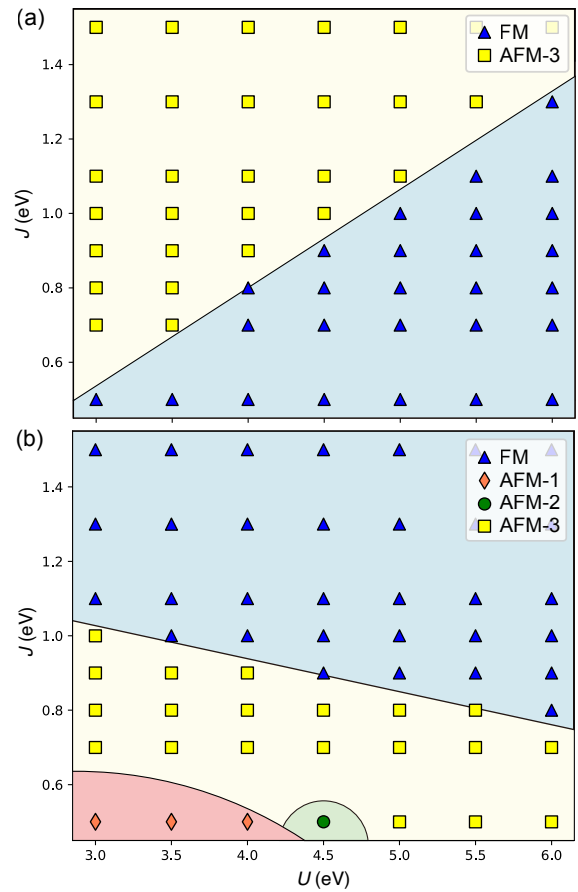

**Fig. 2.** Magnetic phase diagrams of LaMn$_7$O$_{12}$ as functions of the Hubbard $U$ and $J$ parameters, calculated using (a) SDFT+$U$ and (b) CDFT+$U$.

Figures 2(a) and (b) show the magnetic phase diagrams depending on $U$ and $J$, obtained from SDFT+$U$ and CDFT+$U$ calculations, respectively. In both approaches, the theoretical magnetic ground state strongly depends on the values of $U$ and $J$, which indicates the importance of choosing appropriate $U$ and $J$ parameters to accurately describe the material properties. It is worth noting that the AFM-3 phase, the recently reported experimental magnetic structure, appears as the ground state within a specific range of $U$ and $J$ in both SDFT+$U$ and CDFT+$U$: $U = 3$ eV with $J = 0.7$-$1.0$ eV, $U = 3.5$ eV with $J = 0.7$-$0.9$ eV, and $U = 4$ eV with $J = 0.9$ eV.

**Table 1** Calculated energy differences per formula unit and magnetic moments of the A′-site and the B-site Mn ions for the FM and three AFM configurations at $U = 3.5$ eV and $J = 0.8$ eV.

| SDFT+$U$ | | | |
|---|---|---|---|
| **Magnetic structure** | **ΔE (eV)** | **A′-site Mn ($\mu_B$)** | **B-site Mn ($\mu_B$)** |
| FM | 0.000 | 3.744 | 3.711 |
| AFM-1 | 0.003 | 3.652 | 3.586 |
| AFM-2 | 0.084 | 3.659 | 3.566 |
| AFM-3 | -0.024 | 3.658 | 3.582 |
| CDFT+$U$ | | | |
| **Magnetic structure** | **ΔE (eV)** | **A′-site Mn ($\mu_B$)** | **B-site Mn ($\mu_B$)** |
| FM | 0.000 | 3.754 | 3.692 |
| AFM-1 | -0.189 | 3.610 | 3.526 |
| AFM-2 | -0.079 | 3.627 | 3.501 |
| AFM-3 | -0.250 | 3.615 | 3.515 |

To choose the appropriate $U$ and $J$ values for LaMn$_7$O$_{12}$, we further performed cRPA calculations. The calculated values are $U = 3.36$ eV and $J = 0.69$ eV, resulting in an effective $U$ ($U_{\text{eff}} = U - J$) of 2.67 eV. These are within the stability range of the AFM-3 phase in both approaches and closely match $U = 3.5$ eV and $J = 0.8$ eV ($U_{\text{eff}} = 2.7$ eV) among the AFM-3 stabilized points. Thus, we investigated the electronic and lattice properties using these $U$ and $J$ values. Table 1 summarizes the energy differences among FM, AFM-1, AFM-2, and AFM-3, along with the corresponding magnetic moments of the A′-site and B-site Mn ions at $U = 3.5$ eV and $J = 0.8$ eV. In SDFT+$U$, the FM state exhibits the second-lowest energy, while in CDFT+$U$ the AFM-1 is the second stable configuration. The magnetic moments of Mn ions are greater than 3.5 $\mu_B$, which corresponds to the high-spin state of Mn$^{3+}$ ($3d^4$).

We analyzed in more detail the differences in the phase diagrams between the two approaches.

As shown in Fig. 2(a), the SDFT + $U$ calculation stabilizes the AFM-3 phase at low $U$ and high $J$, whereas the FM phase exhibits the lowest energy at high $U$ and low $J$ values. Although four magnetic structures were considered, only the FM and AFM-3 configurations appear as the ground state depending on $U$ and $J$ in this approach. In contrast, the CDFT+$U$ results in Fig. 2(b) show the FM ground state at high $U$ and high $J$, while the AFM phases emerge at low $U$ and low $J$ values. Notably, all four magnetic configurations are present in the CDFT+$U$ phase diagram. Furthermore, the phase boundaries exhibit distinct behaviors: in SDFT+$U$, the boundary shifts upward with increasing $U$, whereas in CDFT+$U$, it slopes downward with increasing $U$. This trend resembles the magnetic phase diagram observed in LaMnO$_3$, a prototype system of LaMn$_7$O$_{12}$ [39,40].

This contrasting behavior can be attributed to the different treatment of spin and charge densities in SDFT+$U$ and CDFT+$U$. In SDFT+$U$, the spin-polarized exchange-correlation functional inherently includes part of the Hund's coupling, resulting in an implicit $J$-dependence of the spin- and orbital-level splitting. As $J$ increases, the spin splitting between majority and minority channels decreases (Fig. S2(a)), thereby enhancing antiferromagnetic superexchange interactions. In contrast, CDFT+$U$ employs a spin-unpolarized density functional, with Hund's coupling introduced explicitly through the $J$ parameter. In this case, increasing $J$ enlarges the spin splitting between majority and minority channels (Fig. S2(b)), which suppresses antiferromagnetic superexchange and stabilizes ferromagnetic order. Furthermore, as shown in Fig. S3(a), the calculated magnetic moment of Mn ions decreases with increasing $J$ in SDFT+$U$, whereas it increases with $J$ in CDFT+$U$ (Fig. S3(b)). In SDFT+$U$, the role of $J$ is effectively to reduce the Coulomb interactions, $U_{\text{eff}}=U-J$, but in CDFT+$U$, one expects a much active role regarding the Hund's exchange. This opposite trend in the magnetic moments is consistent with the opposite $J$-dependence of spin splitting in SDFT+$U$ and

CDFT+$U$, and is also in agreement with results for LaMnO$_3$ [40].

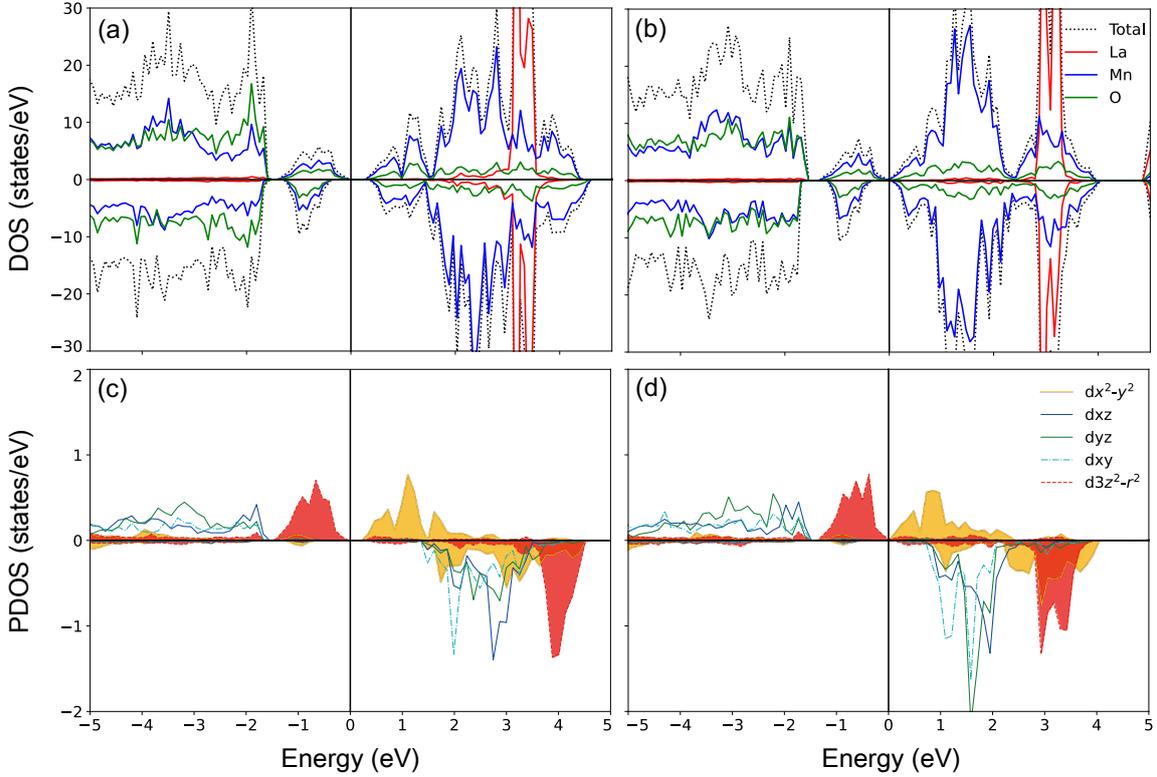

**Fig. 3** Total and atomic density of states (DOS) using (a) SDFT+$U$ and (b) CDFT+$U$. Orbital-projected density of states (PDOS) of B-site Mn-3$d$ orbitals using (c) SDFT+$U$ and (d) CDFT+$U$ ($U$ = 3.5eV, $J$ = 0.8 eV).

Among the magnetic structures of LaMn$_7$O$_{12}$, the AFM-3 configuration, which corresponds to both the experimentally and computationally observed magnetic ground state, was selected for detailed electronic structure analysis. Figure 3 shows the density of states (DOS) of the AFM-3 ground state, calculated at $U$ = 3.5 eV and $J$ = 0.8 eV, selected from the phase diagram (Fig. 2) because it closely matches the cRPA-estimated values. The conduction band minimum and valence band maximum are dominated by Mn-3$d$ states, suggesting that the Mn 3$d$ electrons play a key role in the conductivity of the material. In addition, a significant contribution from O-2$p$ orbitals is observed below the Fermi level, emphasizing the importance of Mn-O hybridization in the valence bands.

As shown in Fig. 3(a), SDFT+$U$ predicts a band gap of 0.23 eV, indicating insulating behavior in agreement with previous experiments [16,17]. In contrast, Fig. 3(b) shows that the CDFT+$U$ calculation at the same $U$ and $J$ values yields a semimetallic state with no band gap. This discrepancy can be attributed to the absence of intrinsic exchange splitting in the charge-only functional, as also observed in LaMnO$_3$, where CDFT+$U$ predicts a smaller band gap than SDFT+$U$ [40]. Consequently, gap formation in CDFT+$U$ depends more strongly on the explicit $U$ value and requires a larger $U$ than SDFT+$U$. When all other conditions are fixed, increasing $U$ to 4.6 eV opens a comparable band gap of 0.22 eV, indicating sufficient electronic localization (Fig. S4).

Figures 3(c) and (d) show the PDOS of the B-site Mn ions, resolved by $d$-orbital character. In both cases, the distinct splitting of the Mn $e_g$ orbitals, namely $\boldsymbol{d_{3z^2-r^2}}$ and $\boldsymbol{d_{x^2-y^2}}$, indicates the presence of the J-T distortions in LaMn$_7$O$_{12}$. The occupied $\boldsymbol{d_{3z^2-r^2}}$ and unoccupied $\boldsymbol{d_{x^2-y^2}}$ states are consistent with the elongated Mn-O bonds along the local octahedral $z$ axis, associated with the J-T distortions of Mn$^{3+}$ ions. The splitting is more pronounced in the SDFT+$U$ calculation, leading to band gap opening and suggesting stronger localization compared to CDFT+$U$. Although the two approaches differ in their treatment of exchange interactions and present variations in the details of DOS, they show qualitatively similar electronic properties of the AFM-3 phase. Both exhibit similar overall DOS shapes, strong Mn-O hybridization near the Fermi level, and a clear splitting of the $e_g$ bands due to the J-T distortions.

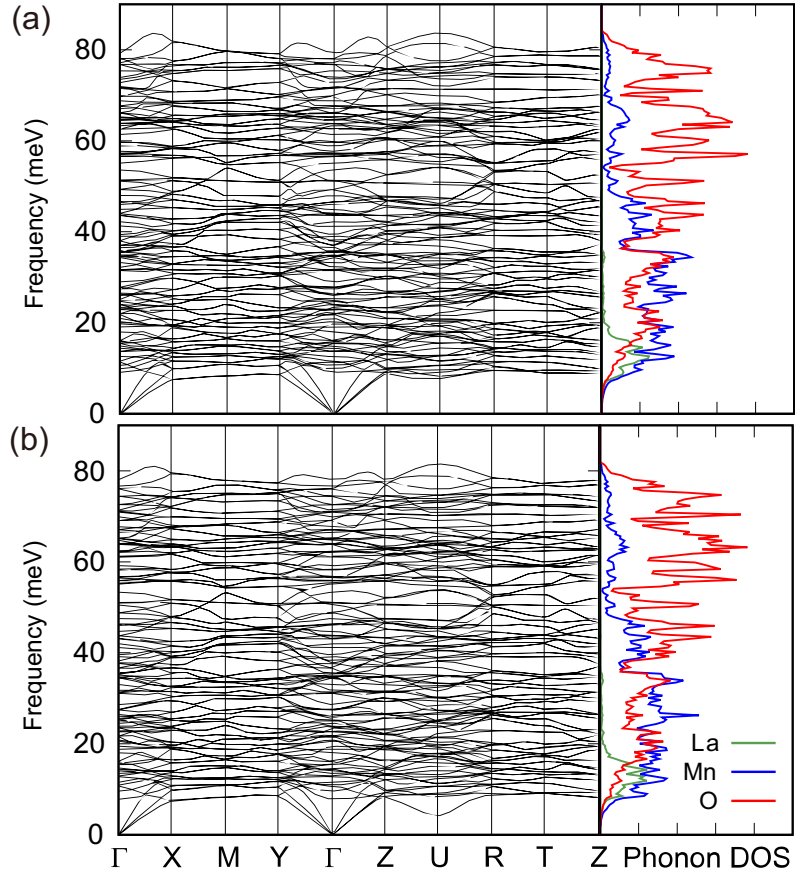

**Fig. 4** Phonon dispersion curves and phonon DOS of the AFM-3 phase, calculated using (a) SDFT+$U$ and (b) CDFT+$U$ with $U = 3.5$ eV and $J = 0.8$ eV.

To further investigate the lattice dynamics, we optimized the structures and calculated phonon dispersion curves using $U = 3.5$ eV and $J = 0.8$ eV. Table 2 summarizes the relaxed structural parameters obtained with SDFT+$U$ and CDFT+$U$, with the relative errors to the experimental values given in parentheses. The calculated lattice parameters show good agreement with experiment, with the relative errors ranging from 0.8 % to 1.7 % in both approaches. The long Mn-O bond length in the J-T distorted $MnO_6$ octahedra is slightly overestimated (3.0% in SDFT+$U$ and 2.0% in CDFT+$U$) but remains within a reasonable range.

**Table 2** Calculated lattice parameters and Mn-O bond lengths in the B-site MnO$_6$ octahedra. The values in parentheses are the relative errors compared to the experiment.

| parameter | SDFT+$U$ | CDFT+$U$ | Exp. [16] |
|---|---|---|---|
| $a$ (Å) | 7.637 (1.7%) | 7.601 (1.2%) | 7.509 |
| $b$ (Å) | 7.421 (1.0%) | 7.410 (0.8%) | 7.349 |
| $c$ (Å) | 7.631 (1.7%) | 7.593 (1.2%) | 7.504 |
| beta (°) | 91.729 (0.4%) | 91.664 (0.3%) | 91.354 |
| long bond (Å) | 2.177 (3.0%) | 2.157 (2.0%) | 2.114 |
| short bond (Å) | 1.907 (0.9%) | 1.907 (0.9%) | 1.890 |

Figure 4 presents the phonon dispersion curves and atom-projected phonon DOS for the AFM-3 structure of LaMn$_7$O$_{12}$ calculated with SDFT+$U$ and CDFT+$U$. In both cases, the absence of imaginary frequencies throughout the Brillouin zone, including the high-symmetry points, demonstrates the dynamical stability of the AFM-3 phase. This stability is consistent with the experimental observation of monoclinic space group *I2/m* structure and AFM-3 magnetic ordering at low temperatures [19].

In the phonon bands, the high-frequency region above 60 meV is dominated by oxygen vibrations, as indicated by the contributions of O atoms in the phonon DOS. This behavior arise from the light atomic mass of oxygen. The intermediate-frequency range (20–60 meV) shows significant contributions from Mn and O atoms, corresponding to vibrational modes involving Mn-O bonds. In the low-frequency range below 20 meV, La atoms contribute the phonon modes due to their large atomic mass. The qualitative similarity of phonon DOS and dispersion between the two methods suggests that both SDFT+$U$ and CDFT+$U$ provide a consistent description of the vibrational properties of the AFM-3 phase, further supporting the robustness of the AFM-3 phase stability.

**Table 3** Calculated Raman-active modes of LaMn$_7$O$_{12}$ compared with experimental Raman frequencies [17].

| No. | SDFT+$U$ | | CDFT+$U$ | | Expt. |
|---|---|---|---|---|---|
| | Sym. | $\omega_0$ (cm$^{-1}$) | Sym. | $\omega_0$ (cm$^{-1}$) | (cm$^{-1}$) [17] |
| 1 | B$_g$ | 163.010 | B$_g$ | 165.686 | |
| 2 | B$_g$ | 199.080 | B$_g$ | 197.750 | 188 |
| 3 | A$_g$ | 200.537 | A$_g$ | 204.303 | 214.12 |
| 4 | B$_g$ | 298.885 | B$_g$ | 304.956 | 290.76 |
| 5 | A$_g$ | 307.113 | A$_g$ | 308.578 | |
| 6 | A$_g$ | 317.823 | A$_g$ | 319.288 | 329.87 |
| 7 | B$_g$ | 351.197 | B$_g$ | 348.050 | |
| 8 | A$_g$ | 398.060 | A$_g$ | 393.753 | |
| 9 | A$_g$ | 413.766 | A$_g$ | 410.685 | 415.35 |
| 10 | A$_g$ | 456.073 | A$_g$ | 446.273 | 431.53 |
| 11 | B$_g$ | 456.795 | B$_g$ | 450.594 | |
| 12 | B$_g$ | 459.637 | B$_g$ | 450.785 | 478.93 |
| 13 | A$_g$ | 533.220 | A$_g$ | 521.762 | |
| 14 | A$_g$ | 563.343 | A$_g$ | 557.366 | 557.31 |
| 15 | B$_g$ | 587.194 | B$_g$ | 563.563 | 575.24 |
| 16 | A$_g$ | 593.087 | B$_g$ | 575.935 | |
| 17 | B$_g$ | 596.240 | A$_g$ | 585.450 | 622.90 |
| 18 | A$_g$ | 628.573 | A$_g$ | 618.079 | 644.21 |

According to group theory, the monoclinic *I2/m* symmetry allows a total of 18 Raman-active modes (10 A$_g$ + 8 B$_g$). Table 3 shows the calculated Raman modes of LaMn$_7$O$_{12}$ with the expected symmetries. Among the 18 Raman-active modes, 11 modes were experimentally observed under ambient conditions [17]. To enable direct comparison, each calculated Raman modes are compared with the nearest experimental frequencies, as listed in Table 3. The deviations between calculated and experimental frequencies are within approximately 6%, indicating good agreement between the computational and experimental results. The consistency with the experiment validates the reliability of both methods in describing the vibrational spectra of LaMn$_7$O$_{12}$.

**Conclusion**

In conclusion, we systematically investigated the magnetic phase diagrams of LaMn$_7$O$_{12}$ as functions of $U$ and $J$ using both SDFT+$U$ and CDFT+$U$ formalisms, considering four magnetic configurations: FM, AFM-1, AFM-2, and AFM-3. Although the magnetic phase diagrams from SDFT+$U$ and CDFT+$U$ show different trends, both approaches identify the AFM-3 ordering as the ground state at $U$ = 3.5 eV and $J$ = 0.8 eV, in agreement with previous experimental observations. These values closely match the cRPA-predicted parameters of $U$ = 3.364 eV and $J$ = 0.692 eV. With these optimal $U$/$J$ values and AFM-3 orderings, the electronic structure exhibits insulating behavior with a band gap of 0.23 eV in SDFT+$U$ and a clear splitting of Mn $e_g$ orbitals associated with the J-T distortions of B-site Mn$^{3+}$ ions. The corresponding phonon dispersion without imaginary frequencies shows the dynamical stability of the AFM-3 structure, and the Raman-active phonon modes are in good agreement with experimental spectra with deviations of less than 6%. This study demonstrates that the appropriate $U$ and $J$ parameters successfully reproduce the experimental magnetic, electronic, and vibrational properties of LaMn$_7$O$_{12}$. These findings provide a basis for further investigations into pressure- or temperature-induced phase transitions and catalytic functionality of LaMn$_7$O$_{12}$.

**Declaration of competing interest**

The authors declare that they have no known competing financial interests or personal relationships that could have appeared to influence the work reported in this paper. The author is an Editorial Board Member/Editor-in-Chief/Associate Editor/Guest Editor for this journal and was not involved in the editorial review or the decision to publish this article.

**Acknowledgements**

This work was supported by the National Research Foundation of Korea (NRF) (Grant No. 2022R1F1A1063011, RS-2025-00557388) and KISTI Supercomputing Center (Project No. KSC-2023-CRE-0395). This work was also supported by Learning & Academic research institution for Master's·PhD students, and Postdocs (LAMP) Program of the NRF grant funded by the Ministry of Education (No. RS-2023-00301914). BK acknowledges support from NRF (NRF2021R1A4A1031920, RS-2021-NR061400, and RS2022-NR068223) and KISTI Supercomputing Center (Project No. KSC-2023-CRE-0413).